\documentstyle[12pt,titlepage]{article}

\font\medio=cmr10 scaled \magstep2
\outer\def\beginsection#1\par{\medbreak\bigskip
      \message{#1}\leftline{\bf#1}\nobreak\medskip
\vskip-\parskip
      \noindent}

\def\laq{\raise 0.4ex\hbox{$<$}\kern -0.8em\lower 0.62
ex\hbox{$\sim$}}
\def\gaq{\raise 0.4ex\hbox{$>$}\kern -0.7em\lower 0.62
ex\hbox{$\sim$}}

\def\beq{\begin{equation}}
\def\eeq{\end{equation}}
\def\bea{\begin{eqnarray}}
\def\eea{\end{eqnarray}}

\def \pa {\partial}
\def \ra {\rightarrow}

\def \fb {\overline \phi}

\def \la {\lambda}
\def \ls {\lambda_s}
\def \La {\Lambda}

\def \b {\beta}

\def \da {\delta}

\def \Om {\Omega}

\def \pfb {\Pi_{\fb}}

\def \pbe {\Pi_{\b}}

\begin{document}
\bibliographystyle {unsrt}

\titlepage
\begin{flushright}
CERN-TH/96-49 \\
hep-th/9602096 \\
\end{flushright}
\vspace{20mm}
\begin{center}
{\bf BIRTH OF THE UNIVERSE AS QUANTUM SCATTERING} \\
{\bf IN STRING COSMOLOGY}

\vspace{10mm}

M. Gasperini\footnote{Permanent address: 
{Dipartimento di Fisica Teorica, Via P. Giuria 1, 10125 Turin,
Italy.}} and G. Veneziano\\
{\em Theory Division, CERN, CH-1211 Geneva 23, Switzerland} \\
\end{center}
\vspace{10mm}
\centerline{\medio  Abstract}

\noindent
In a Wheeler-De Witt approach to quantum string cosmology, the
present state of the Universe arises from the scattering and
reflection of the wave function representing the initial string
vacuum in superspace. This scenario is described and 
compared with the more
conventional quantum cosmology picture, in which the birth of the
Universe is represented as a process of tunnelling ``from nothing"
in superspace.
\vspace{20mm}
\begin{center}
{\it Essay written for the 1996 Awards of the Gravity Research
Foundation} \\
{\it (Wellesley Hills, Ma, 02181-0004)}
\end{center}

\vspace{5mm}

\vfill
\begin{flushleft}
CERN-TH/96-49\\
February 1996 
\end{flushleft}

\newpage

In the standard cosmological model, the birth of 
our Universe is
assumed to coincide with the initial big-bang singularity, 
characterizing
the classical solutions of the Einstein cosmological
 equations. Near the
singularity, however, the Universe approaches the 
Planck curvature
scale and the quantum gravity regime, where a classical
description of the space-time manifold is no longer
appropriate. By adopting, in that regime, a quantum
cosmology approach, one can describe the birth of the
Universe as a ``tunnelling from nothing" \cite{1}-\cite{3},
where the process of tunnelling refers to the
Wheeler-De-Witt (WDW) wave function \cite{4} in
superspace.

In string cosmology models \cite{5}, the Universe starts
evolving  from the string
perturbative vacuum, a state with flat metric and vanishing
coupling constant. The curvature and the coupling grow
during an initial ``pre-big-bang" phase, and this growth,
according to the low-energy effective action, leads
classically to a singular state which marks the beginning of
the standard,  post-big-bang cosmological era. 
By applying a quantum cosmology approach, the transition
through the singular big-bang regime can be described as a 
scattering of the initial pre-big-bang state into a final
post-big-bang configuration, in particular as a reflection of
the WDW wave function in superspace.
The purpose of this paper is to illustrate this effect and to
stress analogies and differences with the more
conventional tunnelling scenario for the birth of the
Universe.

For an easier comparison of the two pictures we shall work
with the simplest example of non-trivial WDW equation, in
which the effective potential is the one induced by a
positive cosmological constant $\Lambda$, due to an
over-critical number of dimensions. We start with the
tree-level, low-energy string effective action \cite{6} \beq
S=-{1\over 2\ls^2}\int d^4x 
\sqrt{-g}e^{-\phi}\left(R+\pa_\mu\phi\pa^\mu\phi
+\Lambda\right) , \label{1}
\eeq
where $\phi$ is the dilaton field, $\ls$ is the
fundamental string-length parameter governing the
derivative expansion of the effective action, and the extra
dimensions have been taken to be completely inert. For a
homogeneous, isotropic and spatially flat
 metric background, with scale factor
$a$ and spatial sections of finite volume, we define:
\beq
\fb= \phi -\ln\int (d^3x/\ls^3) - \sqrt 3 \b  , ~~~~~~~
\b = \sqrt{3}\ln a.
 \label{2}
\eeq
In the cosmic-time gauge, $g_{00} = 1$, the action (\ref{1})
becomes:
\beq 
S=-{\ls\over 2}\int dt 
e^{-\fb}\left(\dot{\fb}^2 - \dot{\beta}^2 + \Lambda 
\right). 
\label{3}
\eeq
By using the convenient time reparameterization $dt =
d\tau e^{-\fb}$ we are finally led to the Hamiltonian
\beq
H= {1\over 2\ls}\left(~\pbe^2-\pfb^2+\ls^2\Lambda
e^{-2\fb}~\right),
\label{4}
\eeq
where $\pbe$, $\pfb$ are the canonical momenta
\beq
\pbe= {\da S \over \da \b '} = \ls \b', 
~~~~~~~~~~ \pfb ={\da S
\over \da \fb '}= -\ls \fb' ,
\label{5}
\eeq
and a prime denotes differentiation with respect to $\tau$.

This Hamiltonian implies momentum conservation along the
$\b$ axis,
\beq
[\pbe ,H] =0, ~~~~~~~~~~~~\pbe = \ls \dot{\b} e^{-\fb} = k =
{\rm const.}
\label{6}
\eeq
The general solution of the classical equations of motion is
well known \cite{7}, \cite{8} and contains two distinct
branches,
\beq
a(t) = a_0 \left(\tanh|\sqrt{\Lambda}t/2|\right)^{\pm
1/\sqrt{3}}, ~~~~~~~\fb = \phi_0 - 
\ln \sinh|\sqrt{\Lambda}t| , ~~~~~ k = \pm \ls
\sqrt{\Lambda} e^{-\phi_0} 
\label{7}
\eeq
where $a_0$ and $\phi_0$ are integration constants. The
two branches are defined over disconnected ranges of time,
$t<0$ and $t>0$, separated by a singularity of the curvature
invariants and of the effective string coupling $e^{\fb}$ at
$t=0$ (for simplicity, we have chosen the integration
constants so as to make the singular ends of both time
ranges coincide at $t=0$).

We are interested, in particular, in the branch describing a classical
approach to the singularity in a state of accelerated expansion,
growing curvature, typical of the pre-big-bang regime \cite{5}
\bea
t<0,~~~~ a&=&a_0\left[\tanh\left(-\sqrt\La
t/2\right)\right]^{-1/\sqrt3}, ~~~~ \fb -\phi_0 = -\ln \sinh
\left(-\sqrt\La t \right) \nonumber \\
\dot a&>&0, ~~~~ \ddot a >0, ~~~~ \dot H >0, ~~~~~~~~
k=\ls\sqrt\La e^{-\phi_0}>0 , 
\label{8}
\eea
and in the branch emerging from the singularity in a state of
decelerated expansion, decreasing curvature,
\bea
t>0,~~~~ a&=&a_0\left[\tanh\left(\sqrt\La
t/2\right)\right]^{1/\sqrt3}, ~~~~ \fb -\phi_0 = -\ln \sinh
\left(\sqrt\La t \right) \nonumber \\
\dot a&>&0, ~~~~ \ddot a <0, ~~~~ \dot H <0, ~~~~~~~~
k=\ls\sqrt\La e^{-\phi_0}>0 . 
\label{9}
\eea
Both branches have positive canonical momentum $\pbe =k >0$,
and are related by a duality transformation including
time-reflection \cite{5,8}, $a(t)\ra a^{-1}(-t)$, $\fb(t) \ra \fb(-t)$.
Let us call, respectively, $(+)$ and $(-)$ the pre- and post-big-bang
branches (\ref{8}) and (\ref{9}). In the high-curvature, strong
coupling regime $\fb \ra +\infty$ (i.e. near the singularity) they are
characterized by a constant and opposite value of the canonical
momentum along $\fb$, namely
\beq
\lim_{\fb \ra +\infty} \pfb^{(\pm)} =
\lim_{\fb \ra +\infty}\left(-\ls \dot{\fb} e^{-\fb}\right)_{\pm}=
\mp k .
\label{10}
\eeq
In the low-energy limit, $\fb \ra -\infty$, the two branches still
have opposite canonical momentum $\pfb$. The momentum is no
longer constant, however, but controlled by $\fb$,
\beq
\lim_{\fb \ra -\infty} \pfb^{(\pm)}  \sim \mp\ls\sqrt\La e^{-\fb}
\label{11}
\eeq

We shall now apply the WDW equation \cite{4}, $H\Psi =0$, to
compute the (classically forbidden) probability of transition from
one branch to another, assuming in particular as the initial state
the pre-big-bang configuration represented classically by the
solution (\ref{8}). For the Hamiltonian (\ref{4}), the WDW equation
is a simple differential equation in the two-dimensional
minisuperspace parameterized by $\fb$ and $\b$,
\beq
\left(~\pa^2_{\fb} -\pa^2_\b +
\ls^2\La e^{-2\fb}~\right)\Psi(\fb, \b)=0 
\label{12}
\eeq
(there is no problem of operator ordering, as the order is uniquely
fixed by the duality symmetries of the string effective action
\cite{9}). By exploiting the conservation property (\ref{6}) we
impose
\beq
\pbe\Psi_k= i \pa_\b\Psi_k=k\Psi_k, 
\label{13}
\eeq
(note the role of the time-like coordinate assigned to $\b$, which
is monotonically increasing from $-\infty$ to $+\infty$), and eq.
(\ref{12}) can be separated by setting
\beq
\Psi_k(\fb,\b) = \psi_k(\fb) e^{-ik\b} , ~~~~~~~~~~~
\left(~\pa^2_{\fb} +k^2+\ls^2\La e^{-2\fb}~\right)\psi_k(\fb)=0
\label{14}
\eeq

The general solution of eq. (\ref{14}) is a linear combination of
Bessel functions \cite{10}, $J_\nu (z)$ and $J_{-\nu}(z)$, of index
$\nu=ik$ and argument $z=\ls\sqrt\La e^{-\fb}$. In order to fix the
boundary conditions we observe that, in the strong coupling
regime $\fb \ra +\infty$, the effective potential of eq. (\ref{14})
becomes negligible, and the WDW solution can be written in the
plane wave form as
\beq
\Psi_{+\infty}^{(\pm)}(\fb, \b)=\lim_{z\ra 0} J_{\pm ik}(z) e^{-ik\b}
\sim e^{-ik(\b \mp \fb)} .
\label {15}
\eeq
In this limit, right- and left-moving waves along $\fb$ correspond,
respectively, to the pre- and post-big-bang classical
configurations (\ref{8}) and (\ref{9}). Indeed, 
\beq
\pfb \Psi_{+\infty}^{(\pm)} = i \pa_{\fb}\Psi_{+\infty}^{(\pm)}
=\mp k \Psi_{+\infty}^{(\pm)}
\label{16}
\eeq
which is the quantum analogue of the classical relation (\ref{10})
(the opposite sign with respect to standard conventions, in the
differential representation of $\pfb$, is due to the negative sign
appearing in the definition (\ref{5})). Consistently with the chosen
pre-big-bang initial conditions, we thus impose that there are only
right-moving waves approaching the singularity at $\fb \ra
+\infty$. This is the same as imposing tunnelling boundary
conditions \cite{3}, which select only outgoing modes at the
(singular) boundary of superspace, and uniquely fixes the WDW
wave function as ($N_k$ is a normalization factor)
\beq
\Psi_k (\fb,\b)= N_k J_{-ik}(\ls\sqrt{\La}e^{-\fb}~)~ e^{-ik\b} .
\label{17}
\eeq

In the low-energy limit $\fb \ra -\infty$, $z \ra \infty$, the effective
potential induced by $\La$ becomes dominant, but the solution can
still be separated into a left- and a right-moving part according to
the asymptotic behaviour of $J_{-\nu}(z)$, namely \cite{10}
\bea
\lim_{\fb \ra -\infty} \Psi_k (\fb,\b) &\sim& 
\Psi_{-\infty}^{(+)} +\Psi_{-\infty}^{(-)} ,
\nonumber \\
\Psi_{-\infty}^{(\pm)}(z, \b) &=& N_k\left(1\over 2 \pi
z\right)^{1/2} \exp \left[-i(k\b \pm z) \pm {i\pi \over 4} \pm {k\pi
\over 2} \right], ~~~z=\ls\sqrt\La e^{-\fb}
\label{18}
\eea
By applying the momentum operator we find
\beq
\lim_{\fb \ra -\infty}\pfb \Psi_{-\infty}^{(\pm)}(\fb,\b)
=\mp z \Psi_{-\infty}^{(\pm)}(\fb,\b)
\eeq
so that $\Psi_{-\infty}^{(+)}(\fb,\b)$ and
$\Psi_{-\infty}^{(-)}(\fb,\b)$ correspond, respectively, to the
pre- and post-big-bang branches of the low-energy classical
solution, according to eq. (\ref{11}). Starting from a pre-big-bang
initial state, we thus obtain a finite probability of transition to the
``dual" low-energy state. The transition is represented as a
reflection of the wave function in minisuperspace \cite{9}, and the
probability is measured by the reflection coefficient $R_k$ as
\beq
R_k={|\Psi_{-\infty}^{(-)}|^2\over |\Psi_{-\infty}^{(+)}|^2}=
 e^{-2\pi k} 
\label{20}
\eeq

By recalling the definition of $k$ and $\fb$, the transition
probability for a three-dimensional portion of space of initial
proper volume $\Om_i$ at $t \ra -\infty$ can be written as
\beq
R(\La, \Om_i, g_s)=\exp\left\{-{\sqrt{12}\pi \over
g_s^2}{\Om_i \over  \ls^3}\left[{\sqrt3\over \ls \sqrt\La} +
\left(1+{3\over \ls^2\La}\right)^{1/2}\right]^{\sqrt3}\right\} ,
\label{21}
\eeq
where $g_s=e^{\phi_s/2}$ is the value of the string coupling at the
scale $H_s= \ls^{-1}$. Quite interestingly, this probability is
independent of $\La$ for $\La >> \ls^{-2}$, and it is peaked in the
strong coupling regime with a typical instanton-like dependence on
the coupling constant, $R \sim \exp(-g_s^{-2})$. This probability is
also invariant under the T-duality transformation \cite{11} 
$(\Om_i/\ls^3) \ra (\ls^3/\Om_i)$, $g_s^2 \ra g_s^2
(\Om_i/\ls^3)^{-2}$. 

By contrast, the probability of birth of the Universe from quantum
tunnelling may be computed from the Einstein-de Sitter action
\beq
S= -{1\over 2 \la_p^2}\int d^4x \sqrt{-g}(R+\La)
\label{22}
\eeq
where $\la_p$ is the Planck length. Solving the corresponding WDW
equation, with appropriate boundary conditions, the tunnelling
probability can be estimated as \cite{1,2,3}
\beq
P \sim \exp \left\{-{4\over \la_p^2 \La}\right\}
\label {23}
\eeq

 The main difference between the above string scenario and the
standard one is that, in the latter, the Universe emerges from the
quantum era in a classical inflationary regime, and the tunnelling
process is completely controlled by the value of the cosmological
constant. By contrast, in our case, the quantum era is
approached at the end of a classical inflationary epoch.
 In both
cases, however, the appropriate boundary conditions imposed at
the big-bang singularity play a crucial role, and the probability of
the birth of our Universe is given by the ratio of the (squared
modulus of the) final wave function representing a
standard decelerated expansion and  the initial wave
function. Also, in both cases the quantum process seems to favour large
values (in Planck or string units) of the cosmological constant.

It is perhaps worth recalling, at this point, 
the words of Vilenkin while  
presenting his ``tunnelling from nothing" scenario \cite{1}-\cite{3}:

{\it ``\dots Here ``nothing" means the vacuum of some more
fundamental theory...."}

\noindent
Our work suggests that ``nothing" can be just the
perturbative vacuum of string theory. The different description of
the birth  process (a wave reflection, rather than a tunnelling)
simply originates from the deep differences  between string and
Einstein gravity, which in the string case 
allow for a long, classical (and
inflationary!) pregnancy.

\end{document}